\documentclass[letterpaper, 10 pt, conference]{ieeeconf}  

\IEEEoverridecommandlockouts                              

\title{\LARGE \bf
Self-supervised TransUNet for Ultrasound regional segmentation of the distal radius in children 
}

\author{\makecell{Yuyue Zhou*, Jessica Knight, Banafshe Felfeliyan, Christopher Keen,\\Abhilash Rakkunedeth Hareendranathan, and Jacob L. Jaremko}  
\thanks{*This work was supported by TD Ready Grant and AMII RAP for financial support and the Alberta Emergency Strategic Clinical Network and Alberta Innovates for clinical scanning.}
\thanks{Yuyue Zhou, Jessica Knight, Banafshe Felfeliyan, Abhilash Rakkunedeth Hareendranathan and Jacob L. Jaremko are with the Department of Radiology and Diagnostic Imaging, University of Alberta, Edmonton, Canada
* Corresponding author
        {\tt\small yuyue2@ualberta.ca}}%
\thanks{ Christopher Keen is with the Department of Biomedical Engineering, University of Alberta, Edmonton, Canada}%
}
\usepackage{makecell}
\usepackage{graphicx}
\usepackage{subfigure} 
\begin{document}

\maketitle
\thispagestyle{empty}
\pagestyle{empty}

\begin{abstract}

Supervised deep learning offers great promise to automate analysis of medical images from segmentation to diagnosis. However, their performance highly relies on the quality and quantity of the data annotation. Meanwhile, curating large annotated datasets for medical images requires a high level of expertise, which is time-consuming and expensive. Recently, to quench the thirst for large data sets with high-quality annotation, self-supervised learning (SSL) methods using unlabeled domain-specific data, have attracted attention. Therefore, designing an SSL method that relies on minimal quantities of labeled data has far-reaching significance in medical images. This paper investigates the feasibility of deploying the Masked Autoencoder for SSL (SSL-MAE) of TransUNet, for segmenting bony regions from children's wrist ultrasound scans. We found that changing the embedding and loss function in SSL-MAE can produce better downstream results compared to the original SSL-MAE. In addition, we determined that only pretraining TransUNet embedding and encoder with SSL-MAE does not work as well as TransUNet without SSL-MAE pretraining on downstream segmentation tasks. 
\end{abstract}

\section{INTRODUCTION}
\subsection{Wrist fracture and ultrasound}
Distal radius fractures are one of the most common fractures in the Emergency Department and account for 8\%-15\% of adult injuries \cite{pogue1990effects}. As children have more pliable bones \cite{randsborg2009distal}, the ratio of children with wrist bone fracture is higher than adults, accounting for up to one-third of all pediatric fractures \cite{handoll2018interventions}. Currently, wrist fracture diagnosis mainly relies on radiographs, however, radiation exposure brings additional risks to infants and young children. Ultrasound (US), with the advantages of no radiation, low cost, and high portability compared to radiograph, could be a potential alternative for wrist fracture detection. The detection and classification of wrist distal radius fractures are based on displacements and comminution of bone shape \cite{meena2014fractures}. Therefore, developing an automatic system for accurate bony region segmentation is a critical step in wrist fracture diagnosis. Nevertheless, compared to wrist fracture detection using radiographs, there are few research on wrist fracture detection with US due to its intrinsic characteristics such as speckle noise and blurred boundaries, making it difficult to interpret and segment. 
\subsection{Image segmentation and TransUNet}
 Most classical medical image segmentation models are developed based on “encoder-decoder” architectures like U-Net \cite{ronneberger2015u}, which allows precise segmentation on the pixel level. While the convolutional kernel is good at dealing with adjacent information, it limits the receptive field of an image, making the model bad at capturing long-range information. Global information is particularly crucial for segmentation, where local patches are often labeled based on the global image context. Transformers, initially designed for natural language processing tasks \cite{vaswani2017attention}, utilize a self-attention embedding mechanism to make models pay more attention to regions with consequential information. Global information will always be kept during the training process of Transformers. Some recent publications have formulated image segmentation as a sequence-to-sequence problem and incorporated Transformers into their architectures to leverage contextual information at every stage \cite{strudel2021segmenter,zheng2021rethinking}. Inspired by the U-Net shape and the success of the Transformer-based image classification model ViT \cite{dosovitskiy2020image}, Chen \textit{et al.} \cite{chen2021transunet} proposed a new segmentation model TransUNet, which leverages advantages of both U-Net and Transformers. Similar to U-Net, TransUNet is a two-stage network that consists of an encoder and a decoder. The embedding, constructed from convolutional layers, is used as a feature extractor to generate a feature map from the input image to the ViT-backbone encoder. TransUNet decoder side is analogous to U-Net's decoder. Like U-Net, skip-connections connect features after embedding to decoder directly. The authors stated that TransUNet surpassed the state-of-arts segmentation models including U-Net. 

\subsection{Self-supervised learning and Masked AutoEncoder}
Although deep learning has made a huge contribution to disease detection and organ/tissue segmentation in medical imaging, its application is largely restricted by limited labeled medical imaging datasets. Unlike natural images, medical image labeling requires people with years of medical training, making it costly and difficult to obtain. Hence, finding a method to train models with limited labeled data is a crucial step toward the widespread application of deep learning in medical imaging. Self-supervised learning (SSL) is an increasingly popular pretraining method, in which models learn the internal and underlying features of images by proxy tasks such as positive/negative image pairing \cite{chen2020simple}, gray-scale image colorization \cite{zhangColorfulImageColorization2016a} and inpainting \cite{pathak2016context} during a pretraining stage with unlabeled data. Models are then initialized with pretraining weights and fine-tuned for the downstream tasks on a small number of labeled images. SSL has been used for medical imaging datasets including US \cite{anand2022benchmarking} and histology images \cite{leiby2022attention}. Felfeliyan \textit{et al.} \cite{felfeliyan2022self} applied different distortions to arbitrary areas of unlabeled data and used the improved Mask R-CNN models to predict distortion type and loss information. He \textit{et al.} \cite{he2022masked} proposed Masked Autoencoders (SSL-MAE) for ViT backbone models self-supervised pretraining. They randomly masked part of the images after patch embedding and trained the model to restore the image with encoder-decoder architecture. As the fine-tuning task could be quite different from the pretraining reconstruction, only encoder weights were kept for downstream tasks. SSL-MAE has been shown to be successful in classification and segmentation tasks with ViT and ViT-based Mask R-CNN models. 

In this study, we extended the SSL-MAE application to TransUNet to determine the utility of SSL-MAE for wrist 2D-US bony region segmentation by TransUNet. We also explored modifying the original loss function and embedding of SSL-MAE. Our ultimate goal was to obtain an accurate model for a small labeled training dataset. 
\section{Methods}
\vspace{-1.5mm}
\subsection{Datasets}
Data was collected prospectively at Stollery Children’s Hospital ED with institutional ethics approval. 118 children aged 0-17 with wrist trauma received 2D-US examinations with Lumify probe in 5 locations: dorsal, proximal dorsal, radial, volar, and proximal volar. Due to logistical and technical issues, 7 children received less than 5 views of US exams and 7 children received more than 5 views. A musculoskeletal sonographer with 10 years of experience labeled the bony region for each patient using a freeware ITK-Snap for medical imaging manual segmentation (Fig. \ref{fig3}). We converted the US video into a sequence of single-image scans for our study. Only slices with distinct and clear views which the sonographer felt to be confident of pathology were labeled and used. Dataset was randomly split into training, validation and test sets based on patient ID to avoid data leakage. For SSL-MAE pretraining, as most adjacent US images looked similar, we selected 1 out of every 10 scans from the original training set for image reconstruction. We then selected 10\% of the pretraining data with bony region labeling for segmentation finetuning. All validation and test set images were kept for segmentation model evaluation. Details of dataset information can be found in Table \ref{table:Table1}.

\begin{table}
\centering
\caption{Detailed description of the datasets for SSL pretraining, and segmentation during training, validation, and testing.}
\label{table:Table1}
\vspace{-4mm}
\resizebox{\linewidth}{!}{%
\begin{tabular}{|>{\centering\hspace{0pt}}m{0.313\linewidth}|>{\centering\hspace{0pt}}m{0.177\linewidth}|>{\centering\hspace{0pt}}m{0.15\linewidth}|>{\centering\hspace{0pt}}m{0.177\linewidth}|>{\centering\arraybackslash\hspace{0pt}}m{0.1\linewidth}|} 
\cline{2-5}
\multicolumn{1}{>{\centering\hspace{0pt}}m{0.313\linewidth}|}{} & SSL-MAE                                               & \multicolumn{3}{>{\centering\arraybackslash\hspace{0pt}}m{0.426\linewidth}|}{Segmentation fine-tuning}                                                               \\ 
\cline{2-5}
\multicolumn{1}{>{\centering\hspace{0pt}}m{0.313\linewidth}|}{} & \begin{tabular}[c]{@{}c@{}}Training\\set\end{tabular} & \begin{tabular}[c]{@{}c@{}}Training\\set\end{tabular} & \begin{tabular}[c]{@{}c@{}}Validation\\set\end{tabular} & \begin{tabular}[c]{@{}c@{}}Test\\set\end{tabular}  \\ 
\hline
\# of patients                                                  & 83                                                    & 81                                                    & 17                                                      & 18                                                \\ 
\hline
\# of sweep videos                                              & 415                                                   & 185                                                   & 84                                                      & 89                                                 \\ 
\hline
\# of images                                                    & 1870                                                  & 187                                                   & 4215                                                    & 3822                                               \\
\hline
\end{tabular}
}
 \vspace{-3mm}
\end{table}

\subsection{SSL-MAE Self-supervised pretraining}
\begin{figure*}[h]
	\centering
	\includegraphics[width=0.75\textwidth]{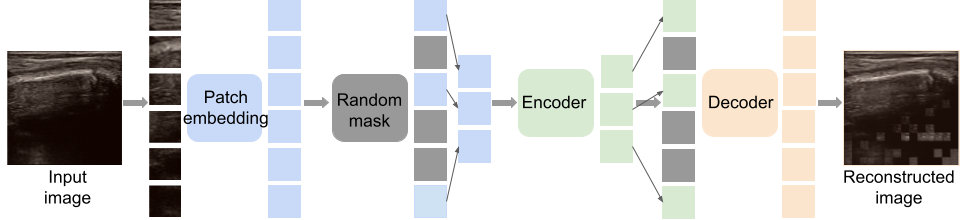}
	  \caption{Masked Autoencoders. ViT patch embedding was changed to ResNet50 patch embedding and ViT encoder was changed to TransUNet encoder.}\label{fig1}
   \vspace{-4mm}
\end{figure*}

The SSL-MAE architecture was utilized for the purpose of self-supervised pretraining. The SSL-MAE pretraining process can be divided into four stages: 1. patch embedding 2. parts of patch embeddings that were masked randomly 3. encoder (input: patch embedding from unmasked region) 4. decoder for image reconstruction (input: latent features generated by encoder+mask tokens) (Fig. \ref{fig1}). To accommodate the TransUNet model, we changed 1. the original SSL-MAE ViT encoder to TransUNet encoder and 2. the SSL-MAE patch embedding to ResNet50 backbone embedding, which is the same as TransUNet embedding. There is a normalization layer between encoder and decoder in the original SSL-MAE, but its impact on the downstream segmentation task was not notable, therefore it was retained for SSL-MAE pretraining. Influenced by Felfeliyan’s work \cite{felfeliyan2022self}, we investigated changing the original \textit{mean squared error (MSE)} loss function to the \textit{root mean squared error (RMSE) + mean absolute error (MAE)} loss over the masked patches reconstruction. Other architectures were maintained the same as the SSL-MAE paper. The encoder and ResNet50 embedding were initialized with ImageNet-pretrained weights for SSL-MAE pretraining.  
\subsection{Segmentation fine-tuning}
Weights of the pretrained models with SSL-MAE were loaded for TransUNet embedding and encoder. TransUNet pretrained on ImageNet was set as a baseline model for comparison. Models were then fine-tuned on an extremely small training set. We used binary-cross-entropy+Dice loss. The model with the smallest validation loss was saved as the best model. We set segmentation threshold based on validation set probability map using Otsu’s method, by minimizing intra-class intensity variance and maximizing inter-class variance.

\subsection{Implementation and Evaluation}
All networks received 3-channel images that were normalized and resized to a size of $224 \times 224$, and were optimized using AdamW optimizer. With hyperparameter tuning, we set the learning rate 0.0001, weight decay 0.05 and batch size 16 for all the experiments. 
 
For self-supervised pretraining task, input images augmented through a random horizontal flip before being forwarded into the model. Mask ratio of 0.75 was selected for pretraining after hyperparameter tuning. All pretraining models were trained for 1200 epochs, and all fine-tuning models were trained for 150 epochs.
A V100 GPU on compute Canada server was used for model training and evaluation. Models were implemented in PyTorch. 1870 unlabeled images were used for pretraining tasks, 187 labeled images were used for training and 4215 and 3822 images were respectively used as validation and test sets.

The cosine similarity was used as the evaluation metrics on SSL-MAE reconstruction over masked area. Dice similarity coefficient (DSC) and Jaccard index were used as the segmentation evaluation metrics. For all  metrics, a higher value implies better performance.

\section{Results}

\subsection{Image reconstruction with SSL-MAE} 
\begin{figure}[h]
	\centering
	\includegraphics[width=0.48\textwidth]{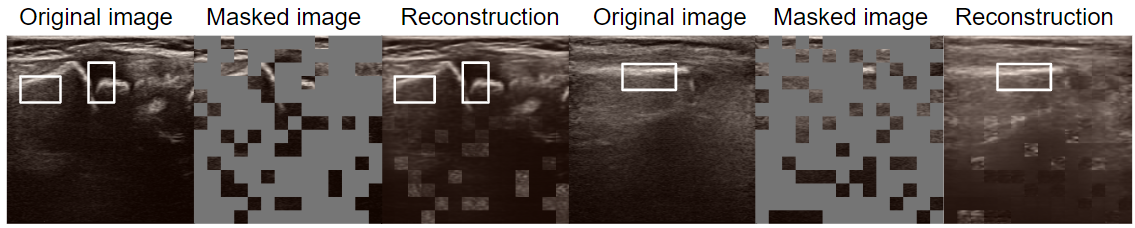}
	  \caption{Masked image reconstruction with SSL-MAE training. The white rectangle shows the comparison between the masked and reconstructed area. }\label{fig2}
   \vspace{-4mm}
\end{figure}

Fig. \ref{fig2} and Table II show the SSL-MAE reconstruction results visually and quantitatively. In general, TransUNet successfully restored the masked region of the original image. The masked area reconstructions by TransUNet are quite close to the masked area on the original image (shown in white rectangles). Changing default loss function to RMSE+MAE loss or changing ViT embedding to ResNet embedding achieved higher cosine similarity between the original image and reconstruction. The image reconstruction results show that the models learned about the global and local information as well as dense image representation during the SSL-MAE pretraining process.

\subsection{Wrist US bony region segmentation}
TransUNet embedding and encoder pretrained with SSL-MAE were then fine-tuned for bony region segmentation. Table II presents quantitative segmentation results on the test set. 
\begin{table*}[h]
\label{table2}
\caption{SSL-MAE reconstruction and wrist US bony region segmentation results (Seg.=Segmentation, CS.=Cosine similarity)}
\begin{center}
\begin{tabular}{|c|c|c|c|c|c|}
\hline
 Model & A &B &C &D &E
\\
\hline
Pretraining& ImageNet &ImageNet+SSL-MAE&ImageNet+SSL-MAE&ImageNet+SSL-MAE&ImageNet+SSL-MAE
\\
\hline
SSL-MAE loss &-&default (MSE)&default (MSE)& RMSE+MAE&RMSE+MAE
\\
\hline
\makecell{SSL-MAE encoder \\embedding}&-&\makecell{default(ViT\\ patch embedding)}&\makecell{ResNet50(TransUNet \\patch embedding)}&\makecell{default(ViT\\ patch embedding)}&\makecell{ResNet50(TransUNet \\patch embedding)}
\\
\hline
\makecell{SSL-MAE reconstruction CS.}&-&0.974&0.982&0.987&\textbf{0.995}
\\
\hline
\makecell{Seg. DSC}&\textbf{0.837}&0.811&0.830&0.831&0.829
\\
\hline
\makecell{Seg. Jaccard index}&\textbf{0.726}&0.690&0.716&0.717&0.714
\\
\hline
\end{tabular}
\end{center}
\end{table*}

TransUNet pretrained on ImageNet was used as the baseline model. The experiment results (Table II) indicate self-supervised pretrained models have very close performance to TransUNet pretrained on ImageNet.
ImageNet pretraining  model achieved the highest DSC (0.837) and Jaccard index. SSL-MAE pretraining led to a low perceptible decrease in model performance (from 0.837 to 0.811-0.831). Out of all experiments, Model B, which is the SSL-MAE + ViT encoder and MSE loss, returned the lowest DSC (0.811). Results show alternating the loss function from MSE loss to RMSE+MAE loss (Model D, E) and replacing SSL-MAE ViT patch embedding with TransUNet ResNet50 patch embedding (Model C, E) improved model B performances but didn't outperform ImageNet pretraining (Model A). 

\begin{figure}[h]
	\centering
	\includegraphics[width=0.43\textwidth]{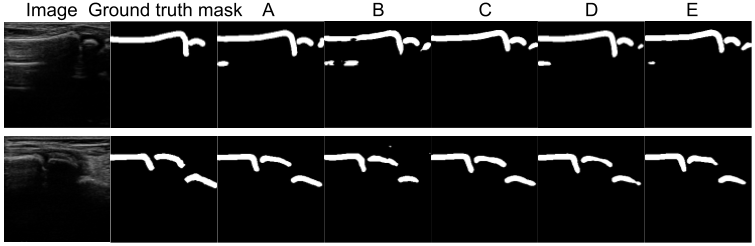}   \vspace{-4mm}
	  \caption{Wrist 2D-US bony region segmentation results. Details of model A-E can be found in Table II.}\label{fig3}
 
\end{figure}

Qualitative results in Fig. \ref{fig3} shows that all models made visually good segmentation. In the upper row image, there is a deceptive artifact 
 and was segmented as a bony region by some models (A, B, D). However, TransUNet with embedding pretrained on SSL-MAE largely shrunk the artifact prediction (C, E). The segmentation image further supports the finding that changing SSL-MAE default ViT patch embedding can help with fine-tuning tasks.

\begin{figure}[h]
	\centering
	\includegraphics[width=0.45\textwidth]{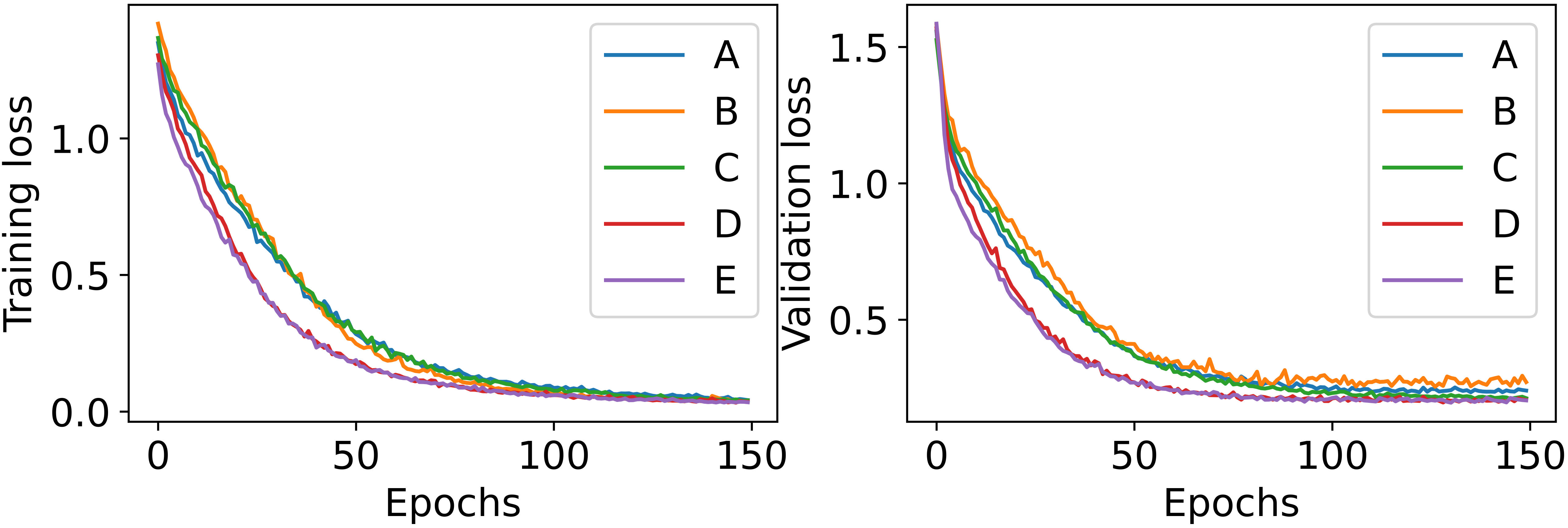}
	  \caption{Segmentation training and validation loss plot}\label{fig4}
   \vspace{-6mm}
\end{figure}

The segmentation training and validation loss plot of all models is demonstrated in Fig. \ref{fig4}. It shows TransUNets pretrained on SSL-MAE with RMSE+MAE loss (Model D and E) converge faster and have lower and more stable validation losses compared to the other models at the beginning(epoch 0-100). The other three models have similar convergence speed.
\vspace{-1mm}

\section{Discussion}
The objective of this study was to determine the advantage of utilizing  SSL-MAE pretraining compared to conventional ImageNet-pretrained TransUNet for wrist US bony region segmentation. The quantitative result shows that SSL-MAE-based pretraining methods achieve close accuracy to the baseline (ImageNet pretraining), but are not able to outperform it. A possible explanation for SSL-MAE TransUNet not surpassing the baseline (TransUNet pretrained with ImageNet) is that only the embedding and encoder weights of SSL-MAE TransUNet were used for the downstream task, and the decoder weights were left out, while skip connections and decoders play a crucial role in the function of TransUNet. Based on results presented in the TransUNet paper \cite{chen2021transunet}, adding skip-connections greatly improved model performance and made TransUNet the best among all other models. However, since SSL-MAE applies random masking after embedding and there is a risk of information leakage from embedding to the decoder, in this study the skip-connections were not used during image reconstruction. Therefore, our next steps will be to include both skip-connections and decoder in a future study with SimMIM self-supervised pretraining \cite{xie2022simmim}, which applies random masking before patch embedding.

Results show that using a combination of two loss functions(RMSE+MAE) for pretraining has a positive effect on convergence speed. This could be attributed to the combination of losses allowing the pretraining task to learn a more variant representation. This suggests that combining multiple loss functions instead of using single MSE loss for SSL-MAE pretraining is beneficial for downstream segmentation tasks, and could help downstream task to converge faster. We will investigate the effect of other loss function combinations in future research.

By comparing cosine similarity from the SSL-MAE reconstruction over masked regions, we can see that higher performance on the pretraining task does not necessarily lead to gain higher performance on the downstream task. It is worth noting that in the SSL-MAE pretraining stage, the unmasked area recoveries were not as optimum as the masked area and had noticeable artifacts, particularly in dark regions. This is because the unmasked regions were not learned by the model during training. Our results are comparable to Xie’s findings in masked SSL \cite{xie2022simmim}.

Currently, the  TransUNet+SSL-MAE has only been pretrained using wrist US scans, for the future we will examine the transferability of features from US scans of other body regions or other modalities.

\vspace{-1mm}
\section{Conclusion}
This study applied Masked AutoEncoder SSL technique to TransUNet pretraining on children's wrists US and fine-tuned TransUNet for bony region segmentation on an extremely small training set with only 187 images. Results showed applying loss function combinations (RMSE+MAE) during SSL-MAE pretraining stage improved the downstream segmentation task compared to using default MSE loss. Pretraining TransUNet patch embedding and encoder did not provide a noticeable improvement in segmentation task.

\addtolength{\textheight}{-12cm}   






\bstctlcite{IEEEexample:BSTcontrol}
\bibliographystyle{IEEEtran}
\bibliography{IEEEabrv,bibliography}

\end{document}